\documentstyle[aps,pre,psfig]{revtex}
\begin{document}
\twocolumn[\hsize\textwidth\columnwidth\hsize\csname@twocolumnfalse\endcsname
\title{About coherent structures in random shell models for passive scalar 
advection}
\author{Luca Biferale$^{1}$, Isabelle Daumont$^{2}$, Thierry Dombre$^{2}$ and 
Alessandra Lanotte$^{3}$\\
\small$^{1}$ Dept. of Physics and INFM, Univ. di Tor Vergata, Via della 
Ricerca Scientifica 1, I-00133 Roma, Italy\\
\small$^{2}$ Centre de Recherche sur les Tr\`es Basses Temp\'eratures-CNRS,\\
Lab. conventionn\'e avec l'Univ. Joseph Fourier,
BP 166, 38042 Grenoble Cedex 9, France\\
\small$^{3}$ CNRS, Observatoire de la C\^ote d'Azur, B.P. 4229,
06304 Nice Cedex 4, France}
\date{\today}
\maketitle
\begin{abstract}
A study 
of anomalous scaling in models of passive scalar advection
 in terms of singular coherent structures is proposed. 
The stochastic dynamical system considered is a shell model reformulation 
of Kraichnan model.  
We extend the  method introduced in \cite{DDG99} to the calculation of 
 self-similar 
instantons and we show how such objects, being the most singular 
events, are appropriate to capture asymptotic scaling properties of the
 scalar field.
Preliminary results concerning the statistical weight of fluctuations around
these optimal configurations are also presented.
\end{abstract}
]

\newcommand{\vect}[1]{ {\bf #1} }
\newcommand{\vectg}[1]{\mbox{\boldmath $#1$}}
\newcommand{\smallvectg}[1]{\mbox{\scriptsize \boldmath$#1$}}
\newcommand{\ra}{\rangle}
\newcommand{\la}{\langle}
\noindent Experimental and numerical 
investigation of passive scalars advected by turbulent
flows have shown that passive scalar structure functions, $T_{2n}(r)$
 have an anomalous power law behaviour\,:
$T_{2n}(r) = 
\la(\theta(x+r)-\theta(x))^{2n}\ra= \la(\delta_r \theta(x))^{2n}\ra
 \sim r^{\zeta(2n)}$, where for anomalous scaling 
we mean that the exponents $\zeta(2n)$ do not follow the dimensional
estimate $\zeta(2n) = n \zeta(2)$.   
A great theoretical challenge is to develop a theory which allows a systematic 
calculation of $\zeta(n)$ from the
Navier-Stokes equations. Recently \cite{KR94}, it has been realized that
intermittent power laws are also present in a model of   
passive scalar advected by stochastic velocity fields, for $n>1$ \cite{GAKU95,CFKL95}.
The model, introduced by Kraichnan, is defined 
 by the standard advection equation:
\begin{equation}
\label{kraichnan}
\partial_t \theta + {\bf u} \cdot \vectg{\partial} \theta = \kappa \Delta
\theta
+\phi, 
\end{equation}
where ${\bf u}$ is a Gaussian, isotropic, white-in-time
stochastic
$d$-dimensional field with a scaling second order structure function:
$\langle(u_i(x)-u_i(x+r))(u_j(x)-u_j(x+r))\rangle =
 D_0 r^{\xi}((d+\xi-1)\delta_{ij}
-\xi r_ir_j/r^2)$. The physical range for the scaling 
parameter of the velocity field is $0 \leq \xi \leq 2$, 
 $\phi$ is an external forcing and $\kappa$ 
is the
molecular diffusivity. \\  
A huge amount of work has been done in the last years on the Kraichnan
model. 
Due to the white-in-time character of the advecting velocity field,
the equation for passive correlators of any order $n$ are
linear and closed. This allows explicit, perturbative  
calculations of anomalous exponents in terms of zero-mode solutions of 
the closed equation satisfied by  $n$-points correlation function,
by means of developments in 
$\xi$ \cite{GAKU95} or in $1/d$ \cite{CFKL95},
 with $d$ the physical space dimensionality.\\
The connection between
anomalous scaling and
zero modes, if fascinating from one side, looks very difficult
to be useful for the most important problem of Navier-Stokes
eqs. In that case, being the problem non linear,
the hierarchy of equations of motion for velocity correlators 
is not closed and the zero-mode approach should be pursued
in a much less handable functional space.\\ From a 
phenomenological point of view, a simple  way 
to understand
the presence of anomalous scaling is to think at 
the scalar field as made of singular scaling fluctuations 
$\delta_r \theta(x) \sim r^{h(x)}$, with a probability to develop
an $h$-fluctuation at scale $r$ given by $P_r(h) \sim r^{f(h)}$, being
$f(h)$ the co-dimension of the fractal set where $h(x) = h$. 
This is the multifractal road 
 to anomalous exponents \cite{Fr95} that leads to the
usual saddle-point estimate for the scaling exponents of structure
functions: $\zeta(2n) = \min_h{(2n h +f(h))}$ \cite{ISM}. 
In this framework, 
high order structure functions are dominated by the most intense events, i.e. 
fluctuation characterized by an exponent  
$h_{min}$: $\lim_{n \rightarrow \infty} \zeta(n) = nh_{min}$.
The emergence of singular fluctuations, at the basis
of the multifractal interpretation, naturally suggests that
instantonic calculus can be used to study
such special configurations in the system.
 Recently, instantons have been successfully applied in the Kraichnan
model to estimate the behaviour of high-order structure functions 
when $d(2-\xi)\gg 1$ \cite{BaLe98}, and to estimate PDF tails 
for $\xi =2$  \cite{FKLM96}.\\
In this letter, we propose   
 an application of the instantonic approach
in random shell models for passive scalar 
advection, where explicit calculation of
the singular coherent structures can be performed.\\ 
Let us briefly summarized our strategy and our main findings.
First, we restrict our hunt for instantons
 to coupled, self-similar, configurations of noise and  passive scalar,
a plausible assumption in view of the multifractal picture
described above. We develop a method for computing in a numerical but exact
way such configurations of optimal Gaussian weight for any scaling exponent
$h$. We find that $h$ cannot go below some finite threshold $h_{min}(\xi)$.
We compare $h_{min}(\xi)$ at varying $\xi$ 
given from the instantonic calculus with those extracted from 
numerical simulation
showing that the agreement is perfect and therefore supporting the idea that
self-similar structures gouvern  high-order intermittency. \\
Second, assuming that these localized pulse-like instantons constitute
the elementary bricks of intermittency also for finite-order moments
we  compute their dressing by quadratic fluctuations.
We obtain in this way the first two terms  of the function $f(h)$
via a ``semi-classical''
expansion. Let us notice that a rigorous
application of the semi-classical analysis would demand for
a small parameter controlling the rate of convergence of the expansion,
like $1/n$ where $n$ is the order of the moment \cite{FKLM96} or $1/d$,
where $d$ is the physical space dimension\cite{BaLe98}.
As we do not dispose of such small parameter in our problem, 
the reliability of our results concerning 
 the statistical weight of the $h$-pulses can only be checked
from an {\it a posteriori}
 comparison with numerical data existing in literature. At the end of
 this communication, we will present some preliminary results on 
such important issue, while much more extensive work will be reported 
elsewhere.\\
\noindent Shell models are simplified dynamical models which have
demonstrated in the past 
to be able to reproduce many of the most important
features of both velocity and passive turbulent cascades \cite{ISM}. \\
The model we are going to use is defined as follows.
First, a shell-discretization
of the Fourier space in a set of wavenumbers defined on a geometric
progression $k_m = k_0 \lambda^m$ is introduced.
Then, passive increments at scale $r_m=k_m^{-1}$
are described by a real variable $\theta_m(t)$. The
time evolution is obtained according to the following criteria\,: (i)
the linear term is purely diffusive and is given by $-\kappa k_m^2 
\theta_m$;
(ii) the advection term is a combination of the form
 $k_m \theta_{m'} u_{m''}$, where $u_m$ are random Gaussian and white-in-time
 shell-velocity fields; (iii) interacting shells are restricted to
nearest-neighbors of $m$; (iv) in the absence of forcing and damping,
 the model conserves the volume in the phase-space and 
the energy $E = \sum_m |\theta_m|^2$. Properties (i), (ii) and 
and (iv) are valid also for the original equation (\ref{kraichnan})
in the Fourier space, while property (iii) is an assumption
of locality of interactions among modes, which is rather well
founded as long as  $0 \ll \xi \ll 2$. 
 The simplest model exhibiting inertial-range intermittency 
is defined by \cite{BBW97}:
\begin{eqnarray}
[\frac{d}{dt} + \kappa k_m^2]\,\theta_m (t) = c_{m}\theta_{m-1}(t)  u_{m}(t) +
 \nonumber  \\
+ a_m \theta_{m-1}(t) u_{m-1}(t) +\delta_{1m} \phi(t),
\label{shellmodel}
\end{eqnarray}
with $a_{m} = -c_{m-1}= k_{m}$, 
and where the forcing term acts only on the first shell. 
Following Kraichnan, we also assume that the forcing term $\phi(t)$ and the
velocity variables $u_m(t)$ are
independent Gaussian and white-in-time random
 variables, with the following scaling prescription for the advecting field:
\begin{equation}
\langle u_m (t) u_n(t') \rangle = \delta(t-t')  k_m^{-\xi} \delta_{mn}.
\end{equation}
Shell models have been proved analytically and non-perturbatively 
\cite{BBW97}
to possess  anomalous zero modes similarly to the original Kraichnan model
(\ref{kraichnan}). \\
The role played by
fluctuations with local exponent $h(x)$ in the original physical 
space model is here replaced by the formation at
 larger scale of structures propagating self-similarly towards 
smaller scales. 
The existence in the inviscid unforced problem of such solutions 
associated with the appearance of finite time singularities is a 
%
The analytical resolution of the instantonic problem 
even in the
 simplified case of shell models is a 
hard task. In \cite{DDG99}, a numerical method to select
self-similar instantons in the case of a shell model for turbulence,
has been introduced. In the following, we are going to apply a similar
method to our case. \\
We rewrite model  (\ref{shellmodel}) in a more
concise form:
\begin{equation}
\label{model}
\frac{d\vectg{\theta}}{dt}= {\rm M}[\vect{b}]\vectg{\theta}.
\end{equation}
The scalar and velocity gradient vectors, $\vectg{\theta}$ and  $\vect{b}$, 
 are made from the
variables $\theta_m$ and $k_mu_m$.
As far as inertial scaling is concerned,  we expect that 
some strong universality 
properties apply with respect to the large scale forcing. Indeed, forcing
changes 
 only the probability with which a pulse appears at large scale,
 but not its inertial range scaling behaviour, $P_{k_m}(h) \sim k_m^{-f(h)}$.
 So, as we are interested only in the evaluation of $f(h)$, we drop the forcing
and dissipation in (\ref{model}). 
The matrix ${\rm M}[\vect{b}]$ is linear in
$\vect{b}$
and can be obviously deduced from (\ref{shellmodel}).
 The stochastic multiplicative equation (\ref{model}) must be 
interpreted {\it \`a la} Stratonovich. 
Nevertheless, once the Ito-prescription for time discretization is adopted,
 the dynamics gets Markovian and a path integral formulation can then be easily
 implemented. This changes (\ref{model}) into:
\begin{equation}
\label{modelito}
 \frac{d\vectg{\theta}}{dt}= -B{\rm D}\vectg{\theta}+ 
{\rm M}[\vect{b}]\vectg{\theta},
\end{equation}
where ${\rm D}$ is a diagonal matrix (Ito-drift) 
${\rm D}_{mm}=k_m^{2-\xi}$, and  $B$ is a positive constant.\\
As we said before, 
we are looking for coherent structures developing a scaling law
$\theta_m \sim k_m^{-h}$ as they propagate towards small scales in
the presence of a velocity realization of optimal Gaussian weight.
 The probability to go from one point to another in configuration
space (spanned by $\vectg{\theta}$) between times $t_i$ and $t_f$
 can be written quite generally as a path 
integral over the three fields $\vect{b}$, $\vectg{\theta}$,
$\vectg{p}$
 of the exponential $e^{-S[\vect{b},
\smallvectg{\theta},\smallvectg{p}]}=e^{-\int_{t_i}^{t_{\!f}}
 {\cal L}[\vect{b},\smallvectg{\theta},\smallvectg{p}]dt}  $, where
 the Lagrangian ${\cal L}$ is
 given by the equation:
\begin{equation}
\label{def_action}
{\cal L}(\vect{b},\vectg{\theta},
\vectg{p})=\frac{1}{2}\vect{b}.{\rm D}^{-1}\vect{b}+
\vectg{p}.(\frac{d\vectg{\theta}}{dt}+B{\rm D}\vectg{\theta}-
 {\rm M}[\vect{b}]\vectg{\theta}), 
\end{equation}
and $\vectg{p}$ is an auxiliary field conjugated to
 $\vectg{\theta}$
which enforces the equation of motion (\ref{modelito}).
 The minimization of the effective action $S$
 leads to the following coupled equations:
\begin{eqnarray}
\frac{d\vectg{\theta}}{dt}&=&-B{\rm D}\vectg{\theta}+ 
{\rm M}[\vect{b}]\vectg{\theta},\label{eqteta}\\
\frac{d\vectg{p}}{dt}&=&B{\rm D}\vectg{p}-\,^{t}{\rm M}[\vect{b}]
\vectg{p},\label{eqzzeta}
\end{eqnarray}
with the self-consistency condition for $\vect{b}$:
\begin{equation} 
\label{eqC}
\vect{b}={\rm D}\,^{t}{\rm N}[\vectg{\theta}]\vectg{p}, 
\end{equation} 
where the matrix ${\rm N}[\vectg{\theta}]$ is defined implicitly through
 the relation 
${\rm N}[\vectg{\theta}]\vect{b}={\rm M}[\vect{b}]\vectg{\theta}$.\\
We are now able to predict the scaling dependence of variables
$b_m$. For a truly self-similar propagation, the cost in action per each
step along the cascade must be  constant.
 The characteristic turn-over time required by
 a pulse localized on the $m-th$ shell to move to
the next one can be dimensionally estimated as $1/(u_m k_m) \equiv b_m^{-1}$.
 Recalling the scaling dependence of ${\rm D}$ and the
definition of action (\ref{def_action}), we expect: $\Delta S =
\int_{t_{m}}^{t_{m+1}} {\cal L}dt \sim
k_m^{-(2-\xi)}b_m$. We can thus deduce that $b_m \sim
k_m^{2-\xi}$. \\
Let us now discuss how to explicitly find solutions of the above
system of equations. Clearly, there is no hope to analytically find 
the exact solutions of these deterministic non linear coupled equations. 
Also numerically, the problem is quite delicate,
 because (\ref{eqteta}) and (\ref{eqzzeta}) are
 obviously dual of each other and have opposite dynamical stability properties.
This phenomenon can be hardly captured by a direct time
integration.
To overcome this obstacle, in \cite{DDG99} it has been proposed a
general 
alternative scheme which adopts an iterative procedure. For a given 
configuration of the noise,
 each step consists in integrating the 
dynamics of the passive scalar (\ref{eqteta}) forward in time to let 
emerge the solution of optimal growth. Conversely, the 
dual dynamics of the auxiliary field (\ref{eqzzeta}) is integrated
backward in time, along the direction of minimal 
growth in agreement with the prediction deduced from (\ref{eqC}): 
$\|\vectg{p}\| 
\sim \|\vectg{\theta}\|^{-1}$. Then the noise $\vect{b}$ can be
 recomputed by the 
self-consistency equation (\ref{eqC}) and the process is repeated 
until the convergence is reached.\\
Self-similar passive solutions must be triggered by self-similar
noise configuration:
\begin{equation}
b_m(t)=\frac{1}{(t^{*}-t)}F(k_m^{2-\xi}(t^{*}-t)),
\label{pippo}
\end{equation}
where $t^{*}$ is the critical time at which a self-similar solution
reaches infinitesimally small scales 
in absence of dissipation. To overcome the non-homogeneity
of time evolution seen by these accelerating pulses,  
we introduce a new time variable $\tau=-\log (t^{*}
-t)$. Then, the advecting self similar velocity field (\ref{pippo})
 can be rewritten under
 the form: $\vect{b(\tau)}=e^{\tau}\vect{C(\tau)}$ where $C_m(\tau )$
 is still
the velocity gradient field, but expressed in a different time scale,
 such that: 
$C_m(\tau )=F(m\,(2-\xi)\log\lambda  -\tau)$.\\ 
The sought self-similar solutions appear in this representation
as traveling waves, whose period  $T=(2-\xi) \log\lambda $ is fixed by
the scaling consideration reported above. In this way, we can limit the
search of solutions on the time interval [$0-T$], and the action
at the final time $t_f=m T$ is deduced by $S({t_f})=m S(T)$.\\
Then comes the main point of our algorithm.
For a fixed noise configuration  $\vect{C}$, the field $\vectg{\theta}$
 must be the eigenvector associated to the maximal
 (in absolute value) Lyapunov exponent $\sigma_{max}$ of the 
Floquet evolution operator:
\begin{equation}
\label{Floquet}
U(T;0)={\cal T}_{-1}\,\exp\,\int_0^{T} 
(-B{\rm D}e^{-\tau}+{\rm M}[\vect{C(\tau)}])d\tau.
\end{equation}
Here ${\cal T}_{-1}$ denotes the translation operator by one unit to
the left along the lattice.
Similarly, the auxiliary field must be the eigenvector associated with 
the Lyapunov exponent $-\sigma_{max}$ of the inverse dual
 operator $^{t}U^{-1}$.\\
Starting from an initial arbitrary traveling wave shape for
$\vect{C}(\tau)$ with  period $T$, we have computed the passive
scalar and its conjugate fields at any time between $0$ and $T$, by
diagonalization of operator $U$, 
recomputed the velocity gradient field $\vect{C(\tau)}$ from the
 self-consistency equation (\ref{eqC}) and iterated this procedure
until an asymptotic stable state,  $\vectg{\theta^{0}}$,
 $\vectg{p^{0}}$, $\vect{C^{0}}$, was reached. 
The scaling exponent $\theta_{m}\sim k_{m}^{-h}$ for the
passive scalar can be deduced by
 $\theta^{0}_{m} (h) 
\sim e^{m \sigma_{max}T}$, so that $h=(\xi-2)\sigma_{max}$.
 Note that $h$ is bound to be positive due to the conservation of energy.
 In our algorithm,
 the norm of the gradient velocity field
$\vect{C}(0)$ acts as the unique control parameter in a one to 
one correspondence with $h$.
The action $S^0(h)$ is, in multifractal language, nothing but the first 
estimate of $f(h)$ 
curve based only on the contribution of all pulse-like solutions, 
more precisely $f(h)=S^0(h)/ln\lambda$.\\
We now turn to the presentation and discussion of our main result.
By varying the control parameter, we obtain a continuum of 
 exponents in the range $ h_{min}(\xi)\leq h \leq h_{max}(\xi)$.
The simple analysis of the $h$-spectrum 
allows predictions only for 
 observable which do not depend on the $f(h)$ curve,
 i.e. only on the scaling of $T_{2n}$ for $n \rightarrow \infty$,
 ($\zeta(n)\sim h_{min}n$ for $n$ large enough).\\ 
Unfortunately, high order exponents are the most difficult quantities to be 
extracted from numerical or experimental data.
 Nevertheless, thanks to the extreme simplicity of shell models, very accurate
numerical simulations have been done \cite{BW96} at different values
of $\xi$ and in some cases a safe upper bound 
prediction on the asymptotic of
$\zeta(n)$ exponents could be extracted.
\begin{figure}
\centerline{\psfig{file=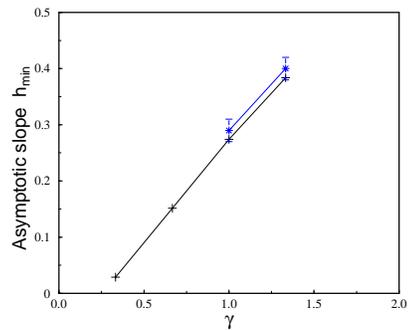,width=6cm}}
\caption{Behavior of the asymptotic slope $h_{min}$ as function of 
 $\gamma=2-\xi$\,: $(+)$ results from the instantonic calculus
for the model of Ref.~\protect\cite{BW96} , $(*)$ upper bound value 
for $h_{min}$ as extracted from numerical integration of the same model.}
\end{figure}
\noindent To compare our results with
 the numerical data  existing in literature,
 we have analyzed the shell-model version 
of passive advection proposed in \cite{BW96}. In Fig.1, we show the 
$h_{min}$ curve obtained at various $\xi$ from instantonic      
calculation, together with the $h_{min}^{num}$ values extracted from
 direct numerical simulation of the quoted model \cite{BW96} performed at 
two different values of $\xi$: the agreement is good. Our 
calculation predicts, within numerical errors,
 the  existence of a critical $\xi_c \sim 1.75 $ 
above  which the minimal exponent reaches the lowest bound $h_{min}=0$.\\
This goes under the name of saturation and it is the 
signature of the presence of discontinuous-like solutions
in the physical space $\delta_r \theta \sim r^0$. Theoretical 
  \cite{BaLe98}
and numerical \cite{FrMaVer98} results suggest the
existence of such effect in the Kraichnan
model for any value of $\xi$.
The existence of saturation in this last is due to typical real-space effects and therefore
it is not surprising that there in not a complete quantitative
analogy with the shell-model case. \\
Let us now present the other -preliminary- result, i.e. 
the role played by instantons for finite-order structure functions.  
If we just keep the zero-th order approximation for 
$f(h)=S_0(h)/\log{\lambda}$, 
 we get the $\zeta_n$ curve shown in Fig.2, 
which is quite far from the numerical results of \cite{BWmm} 
(the asymptotic linear behavior is in fact not even reached
 in the range of $n$
 represented on the figure).
 In order to get a better assessment of the true statistical weight of the 
optimal solutions, we computed the next to leading order term 
in a ``semi-classical'' expansion. Fluctuations around the
 action were developed to quadratic order with
 respect to $\vectg{\theta^{0}}$,
 $\vectg{p^{0}}$, $\vect{C^{0}}$, and the summation over all 
perturbed trajectories leading to the same effective scaling exponent for the 
$\vectg{\theta}$ field after $m$ cascade steps was performed. It 
 turns out (see Fig.2) that the contribution in the action
 of quadratic fluctuations, $S_1(h)$,
greatly improves the evaluation of $\zeta(n)$.
\begin{figure}
\centerline{\psfig{file=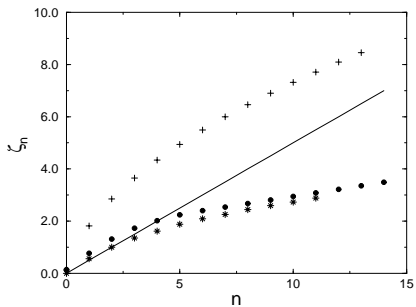,width=6cm}}
\caption{Exponents $\zeta_n$ of structure functions of order $n$ {\it vs} $n$
 for $\xi=1$\,: ($*$) data from direct numerical simulation of Ref.~\protect\cite{BWmm}; $(+)$ Legendre transform of the action $f(h) = S_0$; ($\bullet$) Legendre transform 
of the action $f(h)=S_0+S_1$. The solid line corresponds
to the dimensional scaling.}
\end{figure}
\noindent Naturally, in the absence of any small parameter in the problem, we
cannot take for granted that the next correction(s) 
would not spoil this rather nice agreement with numerical
data. But the surprising fact that $S_0+S_1$ is strongly reduced with
respect to $S_0$, even for the most intense events, does not imply by
itself a lack of consistency of our computation. In any case, the
prediction of the asymptotic slope of the $\zeta_n$ curve, based on the
value $h_{min}$ is obviously valid beyond all orders of 
perturbation. \\Moreover, for values of $\xi>1$, we 
find that the second order exponent extracted from
our calculation is in good agreement the exact result $\zeta_2=2-\xi$, 
suggesting that our approach is able to give relevant statistical 
information also on not too intense fluctuations.\\
\noindent
In conclusion, we have presented an application of the
semi-classical approach in the framework of shell models
for random advection of passive scalar fields.
Instantons are calculated through a numerically assisted 
method solving the equations coming from probability extrema: the
algorithm has revealed capable to pick up those configurations
giving the main contributions to high order moments. 
Of course, we are far from having a systematic, under analytical control,
approach to calculate anomalous exponents in this class of models. 
Nevertheless, the encouraging results here presented raise some relevant 
questions which go well beyond the realm of shell-models. 
To quote just one, 
we still lack a full comprehension of the connection between the usual multiplicative-random process and the instantonic approaches to multifractality: in particular, it is not clear what would be the prediction for 
multi-scale and multi-time correlations of the kind discussed in \cite{bbct}
within the instantonic formulation.\\
\noindent It is a pleasure to thank J-L. Gilson and 
P. Muratore-Ginanneschi for many useful discussions 
on the subject. LB has been partially supported by INFM (PRA-TURBO)
and by the EU contract FMRX-CT98-0175. 

\end{document}